\documentclass[12pt,fleqn]{article}
\usepackage{graphicx,cite,amssymb}

\newcommand{\newsection}[1]{\section{#1}}   

\def\be{\begin{equation}}
\def\ee{\end{equation}}
\def\bea{\begin{eqnarray}}
\def\eea{\end{eqnarray}}

\def\bbuildrel#1_#2^#3{\mathrel{\mathop{\kern 0pt#1}\limits_{#2}^{#3}}}
\def\slash#1{\setbox0=\hbox{$#1$}#1\hskip-\wd0\dimen0=5pt\advance
       \dimen0 by-\ht0\advance\dimen0 by\dp0\lower0.5\dimen0\hbox
         to\wd0{\hss\sl/\/\hss}}

\newcommand{\gsim}{\;\rlap{\lower 3.5 pt \hbox{$\mathchar \sim$}} \raise 1pt \hbox {$>$}\;}
\newcommand{\lsim}{\;\rlap{\lower 3.5 pt \hbox{$\mathchar \sim$}} \raise 1pt \hbox {$<$}\;}

\newcommand{\f}{\frac}

\newcommand{\al}{\alpha_{\mathrm s}}

\begin{document}

\begin{titlepage}

\begin{flushright}
TTP17-004\\
IFT-1/2017\\[3cm]
\end{flushright}
\begin{center}
\setlength {\baselineskip}{0.3in} 
{\bf\Large\boldmath Weak Radiative Decays of the $B$ Meson and\\ 
     Bounds on $M_{H^\pm}$ in the Two-Higgs-Doublet Model }\\[15mm]
\setlength {\baselineskip}{0.2in}
{\large  Miko{\l}aj Misiak$^{1,2}$ and~ Matthias Steinhauser$^3$}\\[5mm]
$^1$~{\it Institute of Theoretical Physics, Faculty of Physics, University of Warsaw,\\
                    02-093 Warsaw, Poland.}\\[5mm]
$^2$~{\it Theoretical Physics Department, CERN, CH-1211 Geneva 23, Switzerland.}\\[5mm]
$^3$~{\it Institut f\"ur Theoretische Teilchenphysik, 
          Karlsruhe Institute of Technology (KIT),\\
          76128 Karlsruhe, Germany.}\\[3cm] 
{\bf Abstract}\\[5mm]
\end{center} 
\setlength{\baselineskip}{0.2in} 

In a recent publication~\cite{Belle:2016ufb}, the Belle collaboration updated
their analysis of the inclusive weak radiative $B$-meson decay, including the
full dataset of $(772 \pm 11)\times 10^6~B\bar B$ pairs. Their result for the
branching ratio is now below the Standard Model
prediction~\cite{Misiak:2015xwa,Czakon:2015exa}, though it remains consistent
with it. However, bounds on the charged Higgs boson mass in the
Two-Higgs-Doublet Model get affected in a significant manner. In the so-called
Model-II, the $95\%\,$C.L. lower bound on $M_{H^\pm}$ is now in the 
$570$--$800\,$GeV range, depending quite sensitively on the method applied
for its determination. Our present note is devoted to presenting and
discussing the updated bounds, as well as to clarifying several ambiguities
that one might encounter in evaluating them. One of such ambiguities stems
from the photon energy cutoff choice, which deserves re-consideration in view
of the improved experimental accuracy.

\end{titlepage}

\newsection{Introduction \label{sec:intro}}

In the absence of any new strongly-interacting particles discovered at the
LHC, one observes growing interest in models where kinematically accessible
exotic particles take part in the electroweak (EW) interactions only.  The
simplest of such models are constructed by extending the Higgs sector of the
Standard Model (SM) via introduction of another $SU(2)_{\rm weak}$
doublet. There are several versions of the Two-Higgs-Doublet Model (2HDM) that
differ in the Higgs boson couplings to fermions. They are usually arranged in
such a way that no tree-level Flavour-Changing Neutral Current
interactions arise~\cite{Glashow:1976nt,Abbott:1979dt}. In the so-called
Model-I, fermions receive their masses in the SM-like manner, from Yukawa
couplings to only one of the two Higgs doublets. In Model-II, the Yukawa
couplings are as in the Minimal Supersymmetric Standard Model, 
i.e. one of the doublets called $H_u$ gives masses to the up-type quarks,
while the other doublet, $H_d$, gives masses to both the down-type quarks and
the leptons.\footnote{
Couplings to leptons are irrelevant for our considerations throughout the
paper. Thus, whatever we write about Models I and II, is also true for the
models called ``X'' and ``Y'', respectively -- see Tabs.~1 and 6 of
Ref.~\cite{Akeroyd:2016ymd}.}

Within the 2HDM, the physical spin-zero boson spectrum consists of one
charged scalar $H^{\pm}$, one neutral pseudoscalar $A^0$, and a pair of
scalars, $H^0$ and $h^0$, the latter of which is identified with the recently
discovered SM-like Higgs boson. If the beyond-SM (BSM) scalars become very
heavy ($M_{H^{\pm}}, M_{A^0}, M_{H^0} \sim M \gg m_{h^0}$), they undergo
decoupling, and the model reduces to the SM at scales much smaller than
$M$. Thus, any claims~\cite{Lees:2013uzd}
%
%
concerning exclusion of the 2HDM in its full parameter space imply
claiming exclusion of the SM, too.

As it is well known (see, e.g., Ref.~\cite{Buras:1993xp}), strong constraints
on $M_{H^{\pm}}$ follow from measurements of the inclusive weak radiative
$B$-meson decay branching ratio.  The most precise results come from the Belle
collaboration, especially from their recent analysis based on the full $(772
\pm 11)\times 10^6~B\bar B$ pair dataset~\cite{Belle:2016ufb}. Their updated
result is now below the Standard Model
prediction~\cite{Misiak:2015xwa,Czakon:2015exa}, though it remains consistent
with it. On the other hand, the 2HDM effects in Model-II can only enhance the
decay rate. In consequence, the lower bound on $M_{H^{\pm}}$ in this model
becomes very strong, reaching the range of $570$--$800\,$GeV. At the same
time, the bound becomes very sensitive to the method applied for its
determination. Given the relevance of the considered bound for many popular
BSM models with extended Higgs sectors, a detailed discussion of this issue is
necessary. This is the main purpose of our present paper.

Following Eqs.~(1.1) and (1.2) of Ref.~\cite{Czakon:2015exa}, as well as
Eq.~(9) of Ref.~\cite{Misiak:2015xwa}, we shall use the CP- and
isospin-averaged branching ratios ${\mathcal B}_{s\gamma}$ and ${\mathcal
B}_{d\gamma}$ of the weak radiative decays, normalizing them to the
analogously averaged branching ratio ${\mathcal B}_{c\ell\nu}$ of the
semileptonic decay. The main observable for our considerations will be
the ratio
\be
R_\gamma ~=~ \f{{\mathcal B}_{s\gamma} + {\mathcal B}_{d\gamma}}{{\mathcal B}_{c\ell\nu}}
~\equiv~ \f{{\mathcal B}_{(s+d)\gamma}}{{\mathcal B}_{c\ell\nu}}.
\ee
We prefer to use $R_\gamma$ rather than ${\mathcal B}_{s\gamma}$ for two
reasons. First, the currently most precise experimental results come from
the fully inclusive analyses of Belle~\cite{Belle:2016ufb} and
Babar~\cite{Lees:2012ym} where the actually measured quantity is
${\mathcal B}_{(s+d)\gamma}$. Second, a normalization
to the semileptonic rate removes the main contribution to the parametric
uncertainty of around 1.5\% on the theory side, while it cannot introduce a
larger uncertainty on the experimental side. Our treatment of the experimental
results will be described in Section~\ref{sec:exp}.

Constraints on $M_{H^\pm}$ from observables other than $R_\gamma$ (or
${\mathcal B}_{s\gamma}$) have been reviewed in several recent articles --
see, e.g., Refs.~\cite{Akeroyd:2016ymd,Moretti:2016qcc,Enomoto:2015wbn}. Their common
property in Model-II is that they become relevant either for small or for very
large ratio of the vacuum expectation values $v_u/v_d \equiv \tan\beta$. On
the other hand, $R_\gamma$ provides a bound that cannot be avoided for any
$\tan\beta$, and turns out to be the strongest one in quite a wide range of
$\tan\beta$.

Our paper is organized as follows. In the next section, the basic framework
for analyzing the considered decays is outlined, with extended explanations
concerning the photon energy cutoff issue. Section~\ref{sec:th} is devoted to
recalling the SM prediction for $R_\gamma$, and discussing the size of
possible extra contributions in the 2HDM. In Section~\ref{sec:exp}, we collect
all the available experimental results for ${\mathcal B}_{s\gamma}$ and/or
${\mathcal B}_{(s+d)\gamma}$, calculate their weighted averages in several
ways, and convert them to $R_\gamma$. The resulting bounds on $M_{H^\pm}$ are
derived in Section~\ref{sec:bounds}. We conclude in Section~\ref{sec:concl}.

\newsection{Photon energy cutoff in ${\mathcal B}_{s\gamma}$ and ${\mathcal B}_{d\gamma}$ \label{sec:phen}}

The radiative decays we are interested in proceed dominantly via quark-level
transitions $b \to s\gamma$,~ $b \to d\gamma$, and their CP-conjugates. A
suppression by small Cabibbo-Kobayashi-Maskawa (CKM) angles makes ${\mathcal
B}_{d\gamma}$ about 20 times smaller than ${\mathcal B}_{s\gamma}$.  For
definiteness, we shall discuss ${\mathcal B}_{s\gamma}$ in what follows,
making separate comments on ${\mathcal B}_{d\gamma}$ only at points where
anything beyond a trivial replacement of the quark flavours matters.

Theoretical analyses of rare $B$-meson decays are most conveniently performed
in the framework of an effective theory that arises after decoupling the
$W$-boson, the heavier SM particles, and all the (relevant) BSM particles. We
assume here that the BSM particles being decoupled are much heavier than the
$b$-quark, but their masses are not much above a TeV. In such a case,
the decoupling can be performed in a single step, at a common renormalization
scale $\mu_0 \sim m_t$. In the effective theory below $\mu_0$, the weak
interaction Lagrangian that matters for $b \to s\gamma$ takes the form
\be
{\cal L}_{\rm weak} \sim \sum_i C_i Q_i,
\ee
where $Q_i$ are dimension-five and -six operators of either four-quark type, e.g.,
\be \label{cc}
Q_1  = (\bar{s}_L \gamma_{\mu} T^a c_L) (\bar{c}_L     \gamma^{\mu} T^a b_L), \hspace{2cm}
Q_2  = (\bar{s}_L \gamma_{\mu}     c_L) (\bar{c}_L     \gamma^{\mu}     b_L),
\ee
or dipole type
\be \label{dipole}
Q_7  =  \f{e}{16\pi^2} m_b (\bar{s}_L \sigma^{\mu \nu}     b_R) F_{\mu \nu}, \hspace{2cm}
Q_8  =  \f{g}{16\pi^2} m_b (\bar{s}_L \sigma^{\mu \nu} T^a b_R) G_{\mu \nu}^a.
\ee
A complete list of $Q_i$ that matter in the SM or 2HDM at the Leading
Order\footnote{
For the Next-to-Leading (NLO) EW corrections, several extra four-quark
operators need to be included -- see Eq.~(2) of Ref.~\cite{Gambino:2001au}. In that
paper, such corrections were calculated within the SM. The 2HDM case is still pending.}
(LO) in $\alpha_{\rm em}$ can be found in Eq.~(1.6) of
Ref.~\cite{Czakon:2015exa}. Their Wilson coefficients $C_i(\mu_0)$ are
evaluated perturbatively in $\al$ by matching several effective-theory
Green's functions with those of the SM or 2HDM. Such calculations have now
reached the Next-to-Next-to-Leading Order (NNLO) accuracy in QCD, i.e. $C_i(\mu_0)$
are known up to ${\mathcal O}(\al^2)$. In the dipole operator case,
performing a three-loop matching~\cite{Misiak:2004ew,Hermann:2012fc} was
necessary at the NNLO.

In the next step, the Wilson coefficients are evolved according to their
renormalization group equations down to the scale $\mu_b \sim m_b$, in order
to resum large logarithms of the form \linebreak
$\left( \al \ln(\mu_0^2/\mu_b^2) \right)^n \sim \left( \al \ln(m_t^2/m_b^2) \right)^n$. 
To achieve this at the NNLO level, anomalous dimension matrices up to four
loops~\cite{Czakon:2006ss} had to be determined. At present, all the Wilson
coefficients $C_i(\mu_b)$ are known with a precision that is sufficient for
evaluating $R_\gamma$ at the NNLO in QCD.

While the calculations of $C_i(\mu_b)$ are purely perturbative, one needs to
take nonperturbative effects into account when determining the physical decay
rates. For $\bar B \to X_s \gamma$ (with $\bar B$ denoting either $\bar B^0$
or $B^-$), the decay rate is a sum of the dominant perturbative contribution
and a subdominant nonperturbative one $\delta\Gamma_{\rm nonp}$, i.e.
\be \label{main}
\Gamma(\bar B \to X_s \gamma) ~=~ \Gamma(b \to X_s^p \gamma) ~+~ \delta\Gamma_{\rm nonp},
\ee
where a photon energy cutoff $E_\gamma > E_0$ in the decaying particle rest
frame is imposed on both sides.\footnote{
The rates would be ill-defined without such a cutoff.}
The partonic final state $X_s^p$ is assumed to consist of charmless quarks and
gluons, while the hadronic state $X_s$ is assumed to contain no charmed or
$c\bar c$ hadrons. The latter requirement is in principle stronger than
demanding that $X_s$ as a whole is charmless. However, all the measurements to
date have been performed with $E_0 \geq 1.7\,$GeV, in which case the $c\bar c$
hadrons and/or pairs of charmed hadrons are kinematically forbidden in $X_s$
anyway. There is no experimental restriction on extra photons or lepton pairs
in $X_s$, but their contribution corresponds to very small NLO QED corrections
that are only partly included on the theory side.
%

The nonperturbative contribution $\delta\Gamma_{\rm nonp}$ in Eq.~(\ref{main})
is strongly dependent on $E_0$. For $E_0 = 1.6\,$GeV, it shifts the SM
prediction for ${\mathcal B}_{s\gamma}$ by almost
$+3\%$~\cite{Buchalla:1997ky},\footnote{
Such a central value of the shift corresponds actually to the effect of $N(E_0)$ in
Eq.~(D.4) of Ref.~\cite{Czakon:2015exa} where a normalization to the semileptonic
rate was used, and some of the nonperturbative effects were relegated to the semileptonic
phase-space factor $C$ in Eqs.~(D.2)--(D.3) there.}
while the corresponding uncertainty is estimated at the $\pm 5\%$
level~\cite{Benzke:2010js}. For higher values of $E_0$, theoretical
uncertainties grow (see below), while the experimental ones decrease thanks to
lower background subtraction errors. To resolve this issue, it has become
standard to perform a data-driven extrapolation of the experimental results
down to $E_0 = 1.6\,$GeV, and compare with theory at that point.

\begin{figure}[t]
\begin{center} 
\includegraphics[width=10cm,angle=0]{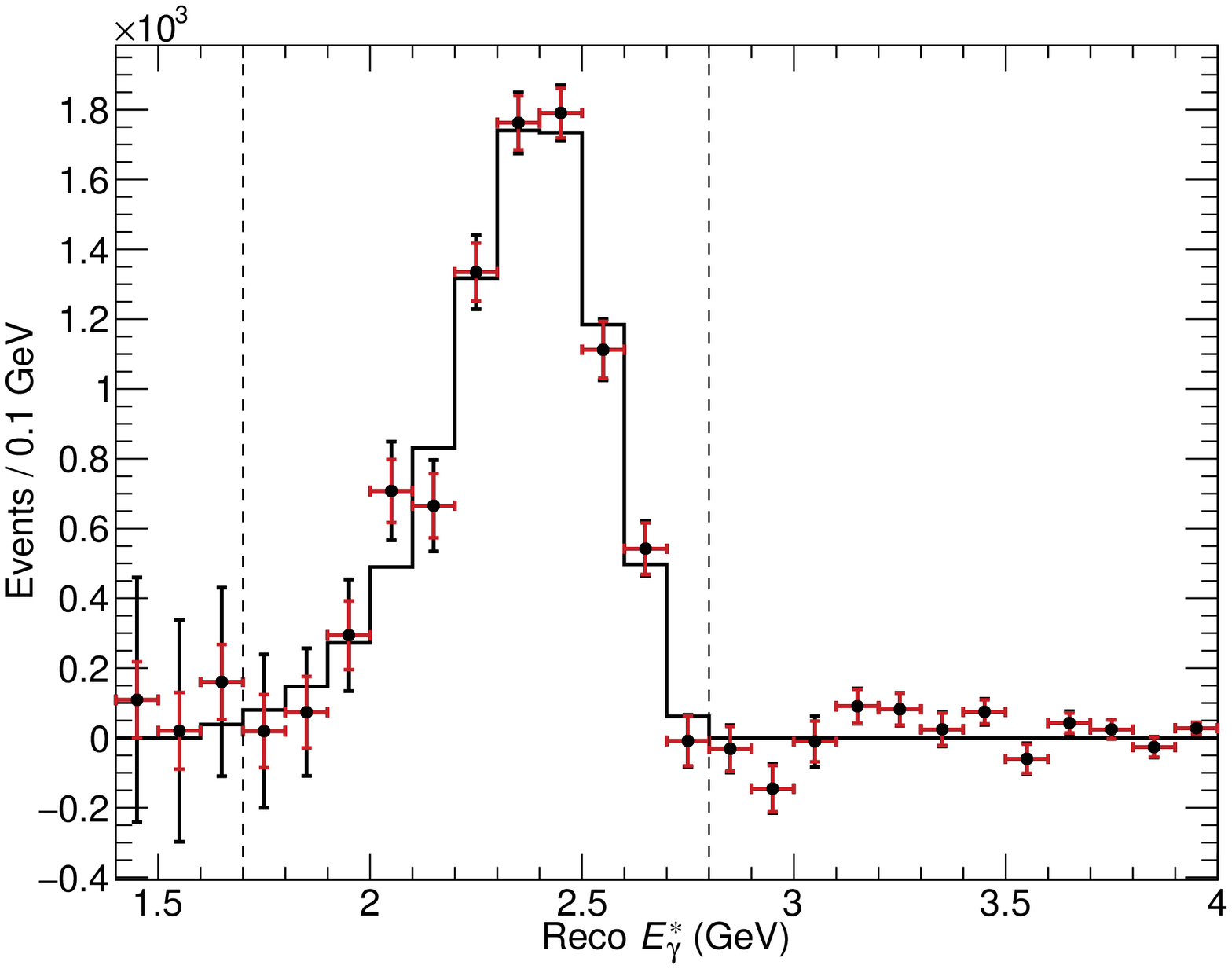}
\caption{\sf Background-subtracted $\bar B \to X_{s+d}\gamma$ photon energy spectrum in the 
$\Upsilon(4S)$ frame, as shown in Fig.~1 of Ref.~\cite{Belle:2016ufb}.
The solid histogram has been obtained by using a shape function model with its parameters
fitted to data.\label{fig:belle}}
\end{center}
\end{figure}

A few comments about such an extrapolation need to be made. First, it is
instructive to have a look at Fig.~\ref{fig:belle} which presents the
background-subtracted photon energy ($E^*_\gamma$) spectrum in the
$\Upsilon(4S)$ frame, as determined by Belle in their full-dataset
measurement~\cite{Belle:2016ufb}.  Photon energies $E_\gamma$ in the $B$-meson
rest frame differ from $E^*_\gamma$ by boost factors that do not exceed 1.07. 
%
%
One can see that energies below $2\,$GeV are well in the tail of the
spectrum. On the other hand, a large set of measurements that gives quite a
precise weighted average for ${\mathcal B}_{(s+d)\gamma}$ is available already
at $E_0 = 1.9\,$GeV (see Section~\ref{sec:exp}). Thus, the extrapolation we
need is really a short one, and only in the tail of the spectrum.

To understand the growth of theoretical uncertainties with $E_0$, one begins
with considering the case when $C_7$ is assumed to be the only nonvanishing
Wilson coefficient at the scale $\mu_b$. In such a case, the fixed-order
Heavy Quark Effective Theory (HQET) formalism can be used to show
that~\cite{Bigi:1992su,Bigi:1992ne,Falk:1993dh}
\be \label{nonp77}
\left[ \f{\delta\Gamma_{\rm nonp}}{\Gamma(b \to X_s^p \gamma)}\right]_{{\rm only}\;C_7} 
~=~ -\f{\mu_\pi^2 + 3\mu_G^2}{2m_b^2} ~+~ 
{\mathcal O}\left(\f{\al \Lambda^2}{(m_b-2E_0)^2}, \f{\Lambda^3}{m_b^3}\right),
\ee
provided~ $m_b - 2 E_0 \gg \Lambda$~ with~ $\Lambda \sim \Lambda_{\rm QCD}$.~ The
quantities $\mu_\pi^2$ and $\mu_G^2$ are of order $\Lambda^2$, and are
currently quite well known from fits to the measured semileptonic decay
spectra~\cite{Alberti:2014yda}.  With growing $E_0$, at some point one enters
into the region where~ $m_b - 2 E_0 \sim \Lambda$,~ and the fixed-order HQET
calculation is no longer applicable. Instead, the leading nonperturbative
effect is parameterized in terms of a universal shape
function~\cite{Neubert:1993um,Bigi:1993ex}. We need to rely on models for this
function, which is the main reason why the theory uncertainties grow with
$E_0$.

A number of shape-function models have been invented in the past, with their
parameters constrained by measurements of the semileptonic and radiative
$B$-meson decay spectra -- see, e.g., Refs.~\cite{Kagan:1998ym,Ligeti:2008ac}.
In Fig.~13 of Ref.~\cite{Ligeti:2008ac}, one can see that the $\bar B \to X_s
\gamma$ photon energy spectrum becomes quite unique already at $E_\gamma =
1.9\,$GeV, at least for the considered class of models.  Such a uniqueness is
indeed expected below the point where the shape-function description starts to
overlap with the fixed-order HQET description. Future studies with precise
Belle-II data should shed more light on the actual location of this point. Our
present approach relies on the assumption that $E_0 = 1.6\,$GeV is definitely
below this point.

Choosing $1.6\,$GeV as the default $E_0$ to compare the fixed-order HQET
predictions with the (extrapolated) experimental results for ${\mathcal
B}_{s\gamma}$ was first suggested in Ref.~\cite{Gambino:2001ew}, at the time
when no precise data on the spectrum were available, and one had to rely on a
limited class of shape-function models.  At present, one might wonder whether
this default $E_0$ might be shifted upwards. However, since such a decision
would need to be made on the basis of the experimental data, shifting the
default $E_0$ could hardly improve anything with respect to the current
extrapolation approach. Another question that one might ask is whether the
extrapolation method (say, from $1.9$ to $1.6$) is indeed superior with
respect to direct measurements at lower values of $E_0$ (but in the range
$[1,6,1.9]$). The answer depends on the balance of uncertainties: the
extrapolation ones and the background subtraction ones. We shall come back to
this issue in Section~\ref{sec:exp}.
\newcommand{\xbf}{\!\mbox{\tiny BF}}    
\newcommand{\xbl}{\!\mbox{\tiny Belle}} 
\newcommand{\xus}{\!\mbox{\tiny fix}}   
\begin{table}[t]
\begin{center}
\begin{tabular}{|c|c|c|c|c|c|}
\hline
&&&&&\\[-4mm]
$E_0\,$[GeV] & $\Delta_s^{\xbf}$ & $\Delta_s^{\xbl}$  & $\Delta_s^{\xus}$ & $\Delta_{s+d}^{\xus}$ & $\Delta_d^{\xus}$ \\[1mm]
\hline
&&&&&\\[-4mm]
$1.7$       & $( 1.5\pm 0.4)\%$ & ?                   &  $1.3\%$          & $1.5\%$               & $5.3\%$           \\[1mm]
$1.8$       & $( 3.4\pm 0.6)\%$ & $(3.69 \pm 1.39)$\% &  $3.0\%$          & $3.4\%$               & $10.5\%$          \\[1mm]
$1.9$       & $( 6.8\pm 1.1)\%$ & ?                   &  $5.5\%$          & $6.0\%$               & $15.7\%$          \\[1mm]
$2.0$       & $(11.9\pm 2.0)\%$ & ?                   &  $10.0\%$         & $10.5\%$              & $22.5\%$          \\
\hline
\end{tabular}
\end{center}
\caption{\sf Quantities $\Delta_q$ from Eq.~(\ref{Deltaq}) evaluated using various approaches.\label{tab:Deltaq}}
\end{table}

Effects of extrapolations from $E_0$ to $1.6\,$GeV can be parameterized by
\be \label{Deltaq}
\Delta_q \equiv \f{{\mathcal B}_{q\gamma}(1.6)}{{\mathcal B}_{q\gamma}(E_0)} - 1,
\ee
with $q=s,d$ or $s+d$. Numerical values of this quantity obtained with the
help of various methods are presented in Tab.~\ref{tab:Deltaq}. Those denoted
by $\Delta_s^{\xbf}$ were evaluated in Ref.~\cite{Buchmuller:2005zv} where
the measured semileptonic and radiative $B$-meson decay spectra (as available
in 2005) were used to determine the $b$-quark mass $m_b$ and the parameter
$\mu_\pi^2$ in three different renormalization schemes. Next, these parameters
were inserted into the Kagan-Neubert shape function model (Eq.~(24) of
Ref.~\cite{Kagan:1998ym}). The shape function was then convoluted with the
perturbatively calculated photon energy spectrum in the $b$-quark decay, which
led to a prediction for the physical photon energy spectrum in the $B$-meson
decay. 

In the next column of Tab.~\ref{tab:Deltaq}, the quantities $\Delta_s^{\xbl}$
were obtained in Ref.~\cite{Belle:2016ufb} using essentially the same method
but with the radiative spectrum only, as measured in the very analysis of
Ref.~\cite{Belle:2016ufb}. In that case, only the result for $E_0 = 1.8\,$GeV
is publicly available at present. The shape function model was used in the
experimental analysis not only for the extrapolation in $E_0$, but also for
efficiency estimates and boosting between $E_\gamma^*$ and $E_\gamma$. The
best fit for $m_b$ and $\mu_\pi^2$ in Ref.~\cite{Belle:2016ufb} leads to a
good description of the measured spectrum (solid histogram in
Fig.~\ref{fig:belle}), and at the same time is consistent with the semileptonic
fits~\cite{Alberti:2014yda}.

The last three columns of Tab.~\ref{tab:Deltaq} have been obtained using the
approach of Refs.~\cite{Misiak:2015xwa,Czakon:2015exa} (perturbative \&
fixed-order HQET), in which case the photon energy spectrum is determined
mainly by the perturbative gluon bremsstrahlung. In these cases, no
uncertainties are quoted, as we do not know at which $E_0$ the fixed-order
HQET description breaks down. The subleading ${\mathcal O}(\al \Lambda^2)$
nonperturbative corrections~\cite{Ewerth:2009yr} begin to rapidly increase at
$E_0$ around $1.8\,$GeV due to $(m_b-2E_0)^2$ in their denominators, but their
overall suppression factor is small, and they remain under control in the
whole region of interest (up to $2\,$GeV).
%

The quantities $\Delta_q^{\xus}$ involve effects of the photon bremsstrahlung
in decays of the $b$ quark to three light (anti)quarks, as calculated in
Refs.~\cite{Kaminski:2012eb,Huber:2014nna}. Such effects are small in
${\mathcal B}_{s\gamma}$ (unless one goes well below $E_0=1.6\,$GeV) but
become much more relevant in ${\mathcal B}_{d\gamma}$ where the tree-level $b
\to du\bar u\gamma$ transitions are not CKM-suppressed with respect to the
leading $b \to d\gamma$ one. In effect, $\Delta_d^{\xus}$ are visibly
different from $\Delta_s^{\xus}$. However, $\Delta_{s+d}^{\xus}$ is not much
different from $\Delta_s^{\xus}$ due to the dominance of ${\mathcal
B}_{s\gamma}$ over ${\mathcal B}_{d\gamma}$. Such photon bremsstrahlung
effects involve collinear singularities in the limit of vanishing quark
masses, which signals the presence of important nonperturbative effects that
need to be described in terms of fragmentation
functions~\cite{Asatrian:2013raa}, and are poorly known. Fortunately, their
overall suppression factors in ${\mathcal B}_{s\gamma}$ and ${\mathcal
B}_{(s+d)\gamma}$ are strong enough, and the corresponding uncertainties are
far below the dominant nonperturbative ones.

It is interesting to observe in Tab.~\ref{tab:Deltaq} that $\Delta_s^{\xbf}$
and $\Delta_s^{\xbl}$ are quite close to $\Delta_s^{\xus}$ and
$\Delta_{s+d}^{\xus}$. It gives us a hope that the breakdown of the
fixed-order HQET description, even if present, is not dramatic in the
considered region of $E_0$. In effect, our sensitivity to ambiguities in
modeling the shape functions is likely to be quite limited, at least for the
purpose of the $1.9 \to 1.6$ extrapolations. However, a devoted analysis with
the most recent data and a wide class of shape function models is necessary to
estimate the corresponding uncertainty in a reliable manner.\footnote{
As follows from Ref.~\cite{Benzke:2010js}, operators other than $Q_7$ give
rise to relevant nonperturbative effects, which may increase the
extrapolation uncertainties.}
Since such an analysis is still awaited, we shall proceed with using
$\Delta_s^{\xbf}$ in what follows for the extrapolation of ${\mathcal B}_{s\gamma}$. 
As far as the extrapolation of ${\mathcal B}_{(s+d)\gamma}$ is concerned, we
are going to rescale $\Delta_s^{\xbf}$ according to the fixed-order results,
namely use $\Delta_{s+d}^{\xbf} \equiv \Delta_s^{\xbf}
\times \Delta_{s+d}^{\xus}/\Delta_s^{\xus}$.

\newsection{The ratio $R_\gamma$ in the SM and 2HDM \label{sec:th}}

Although the perturbative decay rate $\Gamma(b \to X_s^p \gamma)$ in
Eq.~(\ref{main}) may seem straightforward to evaluate, its determination to
better than $\pm 5\%$ accuracy requires including the NNLO QCD corrections,
which is a highly nontrivial task. While the Wilson coefficients are already
known to sufficient accuracy both in the SM and 2HDM (as already mentioned in
the previous section), our knowledge of the NNLO corrections is yet incomplete
in the case of matrix elements, namely interferences among on-shell decay
amplitudes generated at the scale $\mu_b$ by the operators $Q_i$. The matrix
elements are the same in the SM and in the 2HDM.

At the NNLO level, we can restrict our attention to the operators listed in
Eqs.~(\ref{cc})--(\ref{dipole}), as the remaining ones can be neglected due to
their small Wilson coefficients. The $Q_7$-$Q_7$ and $Q_7$-$Q_8$ interference
terms are already known at ${\mathcal O}(\al^2)$ in a complete
manner~\cite{Blokland:2005uk,Melnikov:2005bx,Asatrian:2006rq,Ewerth:2008nv,Asatrian:2010rq}. The
NNLO interference terms not involving $Q_7$ can be separated into two-body
final state contributions (trivially derived from the NLO results) or
relatively small $(n\geq 3)$-body final state contributions that have been
calculated so far~\cite{Ligeti:1999ea,Ferroglia:2010xe,Misiak:2010tk} only in
the Brodsky-Lepage-Mackenzie (BLM)~\cite{Brodsky:1982gc} approximation. The
main perturbative uncertainty comes from the $Q_{1,2}$-$Q_7$ interferences at
${\mathcal O}(\al^2)$. Their BLM parts, as well as effects of nonvanishing
quark masses on the gluon lines were evaluated in
Refs.~\cite{Ligeti:1999ea,Bieri:2003ue,Boughezal:2007ny} for arbitrary values
of the charm quark mass $m_c$.  The remaining parts were found only in the
limits $m_c \gg m_b/2$~\cite{Misiak:2010sk} or $m_c =
0$~\cite{Czakon:2015exa}, and then an interpolation between these two limits
was performed~\cite{Czakon:2015exa}.

With all the NNLO QCD, NLO EW and nonperturbative corrections evaluated
to date, the SM prediction for $R_\gamma$ at $E_0=1.6\,$GeV
reads~\cite{Misiak:2015xwa}
\be \label{rsm}
R_\gamma^{\rm SM} = (3.31 \pm 0.22) \times 10^{-3},
\ee
where the overall uncertainty has been obtained by combining in quadrature the
nonperturbative one $(\pm 5\%)$, the parametric one $(\pm 1.5\%)$, the one
stemming from neglected higher-order effects $(\pm 3\%)$, and the one due to
the above-mentioned interpolation in $m_c$ $(\pm 3\%)$.
\begin{figure}[t]
\begin{center}
\includegraphics[width=8cm,angle=0]{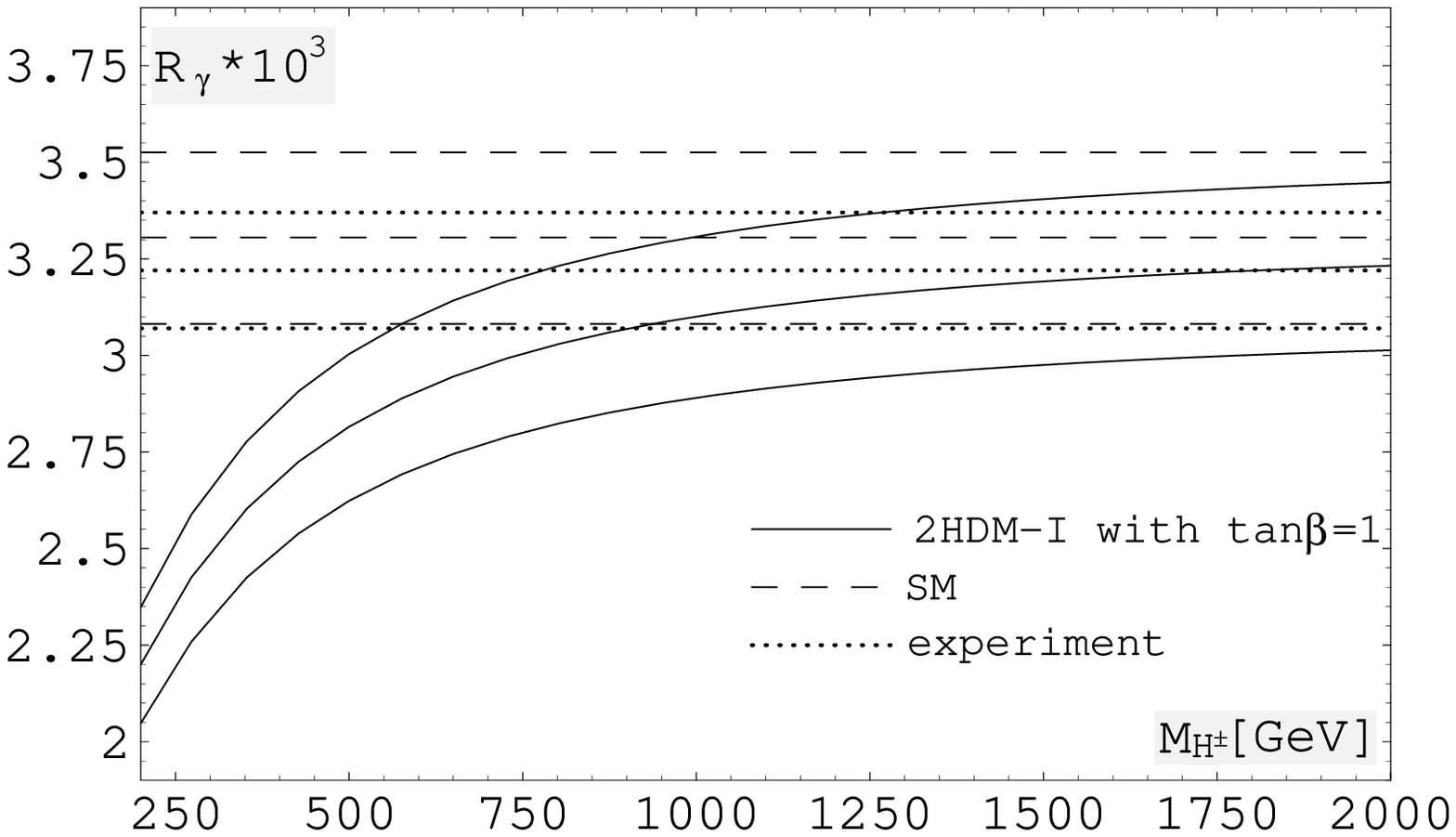}\hspace{5mm}
\includegraphics[width=8cm,angle=0]{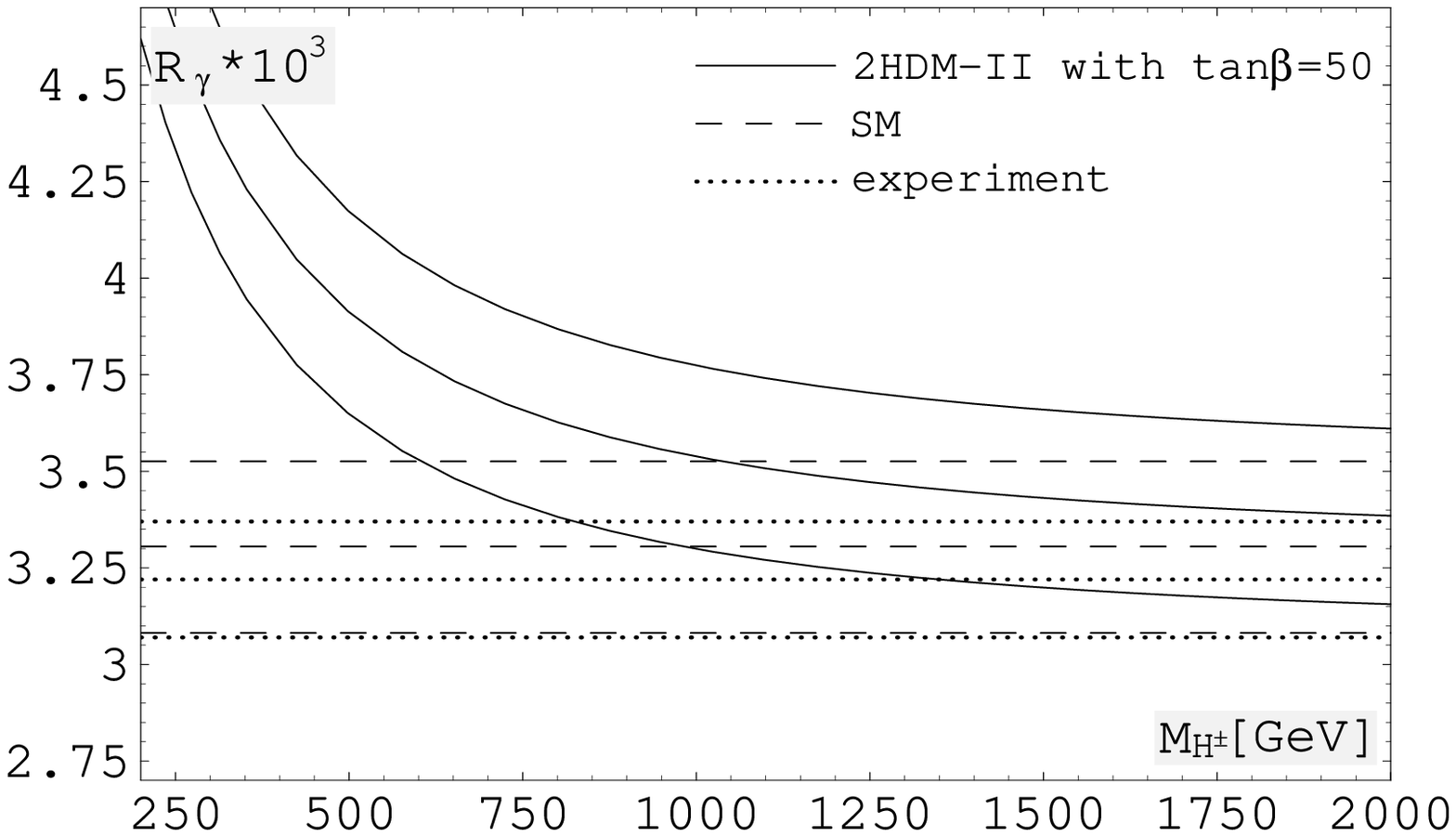}
\caption{\sf $R_\gamma$ at $E_0=1.6\,$GeV as a function of $M_{H^{\pm}}$ in
Model-I with $\tan\beta=1$ (left) and in Model-II with $\tan\beta=50$
(right). Middle lines show the central values, while the upper and lower ones
are shifted by $\pm 1\sigma$. Solid and dashed curves correspond to the 2HDM
and SM predictions, respectively. Dotted lines show the experimental average
$R_{\gamma}^{\exp} = (3.22 \pm 0.15) \times 10^{-3}$ (see
Section~\ref{sec:exp}).\label{fig:MHc}}
\end{center}
\end{figure}

In the 2HDM, additional contributions to the Wilson coefficient matching arise
from diagrams with the physical charged scalar exchanges. The relevant
couplings and sample diagrams can be found, e.g., in Sec.~2.3 of
Ref.~\cite{Hermann:2012fc}. We evaluate $R_\gamma$ in Model-I and Model-II
with the same accuracy as in the SM, up to the missing NLO EW corrections to
the charged Higgs contributions. Apart from the SM parameters, the results
depend only on $M_{H^{\pm}}$ and $\tan\beta$. They are plotted in
Fig.~\ref{fig:MHc} as functions of $M_{H^{\pm}}$ in two cases of particular
interest: Model-I with $\tan\beta=1$ and Model-II with $\tan\beta=50$. The
solid and dashed curves in these plots correspond to the 2HDM and SM cases,
respectively. Dotted lines indicate the experimental average to be discussed
in the next section.

In Model-I, the charged Higgs contribution to the decay amplitude is
proportional to $\cot^2\beta$, and it interferes with the SM one in a
destructive manner.  In Model-II, the interference is always constructive, and
the charged Higgs amplitude has the form\footnote{ 
The term proportional to $\tan^2\beta$ is suppressed by the strange quark
mass, and we neglect it here.}
$A + B \cot^2\beta$. The quantities $A$ and $B$ depend on $M_{H^{\pm}}$ only,
and they have the same sign. In consequence, an absolute bound on
$M_{H^{\pm}}$ can be derived from $R_\gamma$ in Model-II by setting the
$\cot^2\beta$ term to zero.  In practice, $R_\gamma$ begins to be practically
independent of $\tan\beta$ already around $\tan\beta \simeq 2$. The
$\tan\beta=50$ case in Fig.~\ref{fig:MHc} indicates that the absolute bound on
$M_{H^{\pm}}$ is going to be in the few-hundred GeV region.  In Model-I,
sizeable deviations of $R_\gamma$ from its SM value occur only for moderate or
small values of $\tan\beta$. The $\tan\beta=1$ case displayed in Fig.~\ref{fig:MHc} 
shows that our sensitivity to $M_{H^{\pm}}$ in this case is almost as strong as
in Model-II.

\newsection{Determining the current experimental average for $R_\gamma$\label{sec:exp}}
\begin{table}[t]
\newcommand{\w}[1]{          $\!\scriptstyle #1\!\!$}
\newcommand{\y}[1]{\boldmath $\!\scriptstyle #1\!\!$}
\begin{center}
\begin{tabular}{|c|c|c|c|c|c|c|c|c|c|c|c|c|}\hline
& \multicolumn{4}{c|}{Babar} & \multicolumn{3}{c|}{Belle} & $\!\!$CLEO$\!\!$ & w.a. & w.a. & $R_\gamma$ & $R_\gamma$ \\
$E_0$ & \cite{Lees:2012ym}   & \cite{Lees:2012wg} & \cite{Aubert:2007my} & aver 
       & \cite{Belle:2016ufb} & \cite{Saito:2014das} & aver 
       & \cite{Chen:2001fja} & ($E_0$) & (1.6) & ($E_0$) & (1.6) \\\hline
1.7 &             &             &             &             & \w{306(28)} &  
    & \w{306(28)} &             & \w{306(28)} & \w{311(28)} &             &             \\
    &             &             &             &             & \y{320(29)} & 
    & \w{320(29)} &             & \w{320(29)} & \w{326(30)} & \w{300(28)} & \w{305(28)} \\\hline
1.8 & \w{321(34)} &             &             & \w{321(34)} & \w{301(22)} &  
    & \w{301(22)} &             & \w{307(19)} & \w{318(19)} &             &             \\
    & \y{335(35)} &             &             & \w{335(35)} & \y{315(23)} & 
    & \w{315(23)} &             & \w{321(19)} & \w{333(20)} & \w{301(19)} & \w{312(19)} \\\hline
1.9 & \w{300(24)} & \y{329(52)} & \w{366(104)}& \w{308(22)} & \w{294(18)} & \y{351(37)} 
    & \w{305(16)} &             & \w{306(13)} & \w{327(14)} &             &             \\
    & \y{313(25)} & \w{344(54)} & \y{381(108)}& \w{321(23)} & \y{307(19)} & \w{367(39)} 
    & \w{319(17)} &             & \w{320(14)} & \w{343(15)} & \w{300(14)} & \w{322(15)} \\\hline
2.0 & \w{280(19)} &             & \w{339(79)} & \w{283(18)} & \w{279(15)} &  
    & \w{279(15)} & \w{293(46)} & \w{281(11)} & \w{315(14)} &             &             \\
    & \y{292(20)} &             & \y{353(83)} & \w{296(19)} & \y{292(15)} & 
    & \w{292(15)} & \y{306(49)} & \w{294(11)} & \w{331(14)} & \w{276(11)} & \w{310(14)} \\\hline
\end{tabular}
\end{center}
\caption{\sf Experimental results and their averages for ${\mathcal
B}_{s\gamma}\times 10^6$ (upper rows) and ${\mathcal B}_{(s+d)\gamma}\times
10^6$ (lower rows) at each value of $E_0$. Each world average (w.a.) is first
calculated at $E_0$ (10th column), and then extrapolated to $1.6\,$GeV (11th
column) using $\Delta_s^{\xbf}$ or $\Delta_{s+d}^{\xbf}$ from
Section~\ref{sec:phen}. In the last two columns, the ratios $R_\gamma (\times
10^5)$ are calculated from the corresponding averages for ${\mathcal
B}_{(s+d)\gamma}$ using ${\mathcal B}_{c\ell\nu} = 0.1067(16)$. \label{tab:averages}}
\end{table}

All the available measurements of ${\mathcal B}_{(s+d)\gamma}$ and ${\mathcal
B}_{s\gamma}$, as well as our averages of them are collected in
Tab.~\ref{tab:averages}. The results of Babar have been obtained using three
methods: fully inclusive~\cite{Lees:2012ym},
semi-inclusive~\cite{Lees:2012wg}, and the hadronic-tag
one~\cite{Aubert:2007my}. Belle has used the fully
inclusive~\cite{Belle:2016ufb} and semi-inclusive~\cite{Saito:2014das}
approaches, while their hadronic-tag analysis is still awaited. In the
measurement of CLEO~\cite{Chen:2001fja}, the fully inclusive method was used.

The most precise results come from the fully inclusive analyses where the
actually measured quantity is ${\mathcal B}_{(s+d)\gamma}$. The same refers to
the hadronic-tag result of Babar, which is actually also fully inclusive. In
the semi-inclusive cases, a single kaon in the final state was required, so the
measurements accounted directly for ${\mathcal B}_{s\gamma}$. We indicate this
in Tab.~\ref{tab:averages} by typesetting the corresponding numbers in bold.

Belle and CLEO provided their ${\mathcal B}_{(s+d)\gamma}$ results explicitly,
while Babar rescaled them to ${\mathcal B}_{s\gamma}$, quoting in each case
the necessary CKM factor together with its uncertainty. In
Tab.~\ref{tab:averages}, we ``undo'' the rescaling using precisely the same
factors.  On the other hand, in the two semi-inclusive cases, we derive
${\mathcal B}_{(s+d)\gamma}$ from ${\mathcal B}_{s\gamma}$ using a rescaling
factor $(1.047 \pm 0.003)$ that we calculate at $E_0=1.9\,$GeV as in
Refs~\cite{Misiak:2015xwa,Czakon:2015exa}.  Our factor differs only slightly
(by $0.2\%$) from $1+|V_{td}/V_{ts}|^2$, due to the $b \to du\bar u\gamma$
effects. Rescaling the semi-inclusive results is a minor issue anyway, as
they come with considerably larger experimental errors.

The reader is referred to the original experimental
papers~\cite{Belle:2016ufb,Lees:2012ym,Lees:2012wg,Aubert:2007my,Saito:2014das,Chen:2001fja}
for the decomposition of errors into the statistical, systematic and
occasionally the spectrum-modeling ones. Here we have added them in quadrature
for the purpose of determining our naive averages, in which no correlations
have been taken into account.  In several cases, we can compare our averages
with the very recent ones of HFAG~\cite{Amhis:2016xyh} where, we believe, the
necessary correlations have been included. For instance, the two Belle results
for ${\mathcal B}_{s\gamma}$ at $E_0=1.9\,$GeV lead to the naive average of~
${\tt av}[ 294(18), 351(37)] = 305(16)$, which perfectly agrees with
Ref.~\cite{Amhis:2016xyh}. In the same row of Tab.~\ref{tab:averages}, the two
less precise results of Babar give~ ${\tt av}[ 329(52), 366(104)] = 336(46)$,
which again overlaps with Ref.~\cite{Amhis:2016xyh}. In this case, the most
precise result of Babar has not been included in the HFAG average. We have
been informed that this point is going to be corrected soon~\cite{MCpriv}.

As far as the world average for ${\mathcal B}_{s\gamma}$ extrapolated to
$E_0=1.6\,$GeV is concerned, Ref.~\cite{Amhis:2016xyh} gives $(3.32 \pm
0.15)\times 10^{-4}$, which is quite close to our $(3.27 \pm 0.14)\times
10^{-4}$ in the row containing the semi-inclusive measurements. We do not know
which inputs have been used in this average of HFAG. Concerning the
extrapolation, they have indicated using the method of
Ref.~\cite{Buchmuller:2005zv}.
\begin{figure}[t]
\begin{center}
\includegraphics[width=8cm,angle=0]{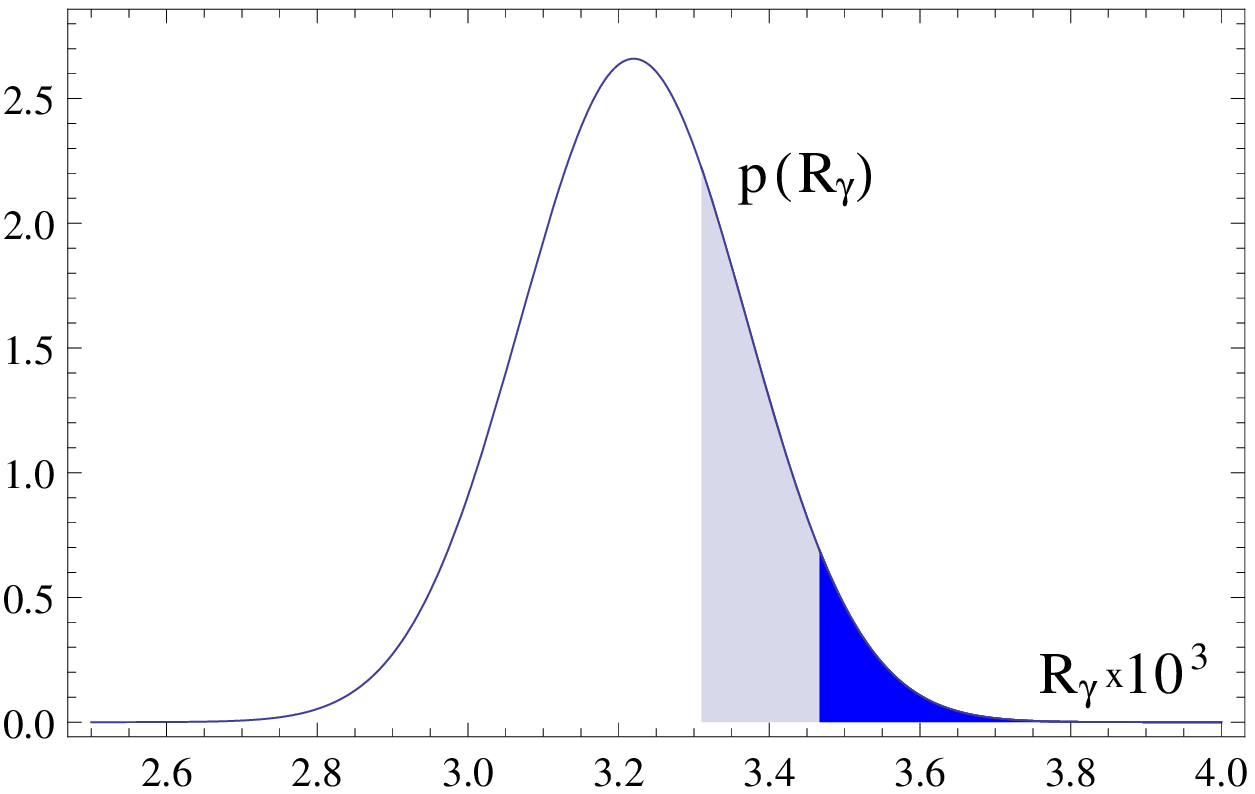}\hspace{5mm}
\includegraphics[width=78mm,angle=0]{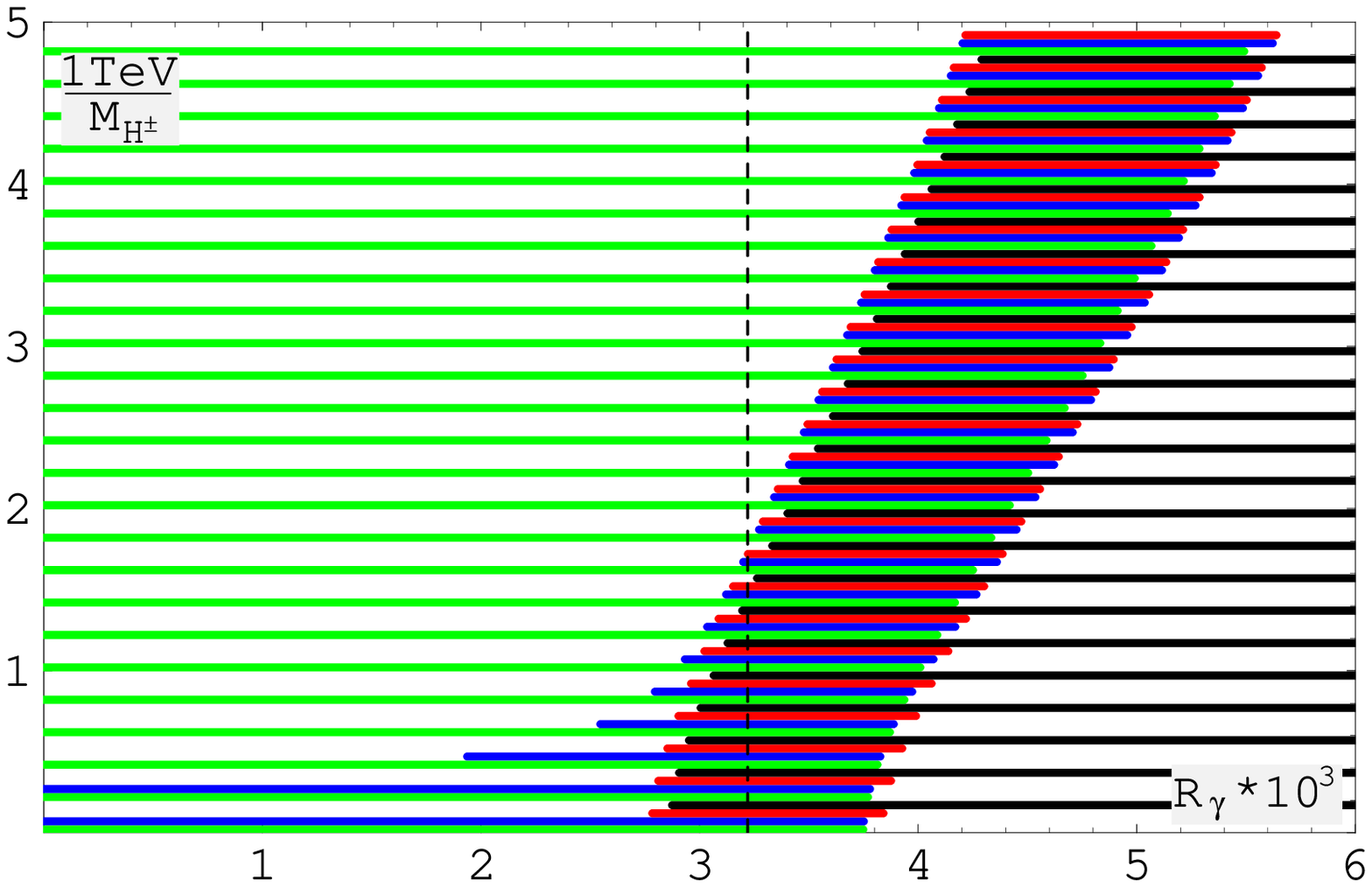}
\caption{\sf Left: Probability density for $R_{\gamma}^{\exp} = (3.22 \pm
             0.15) \times 10^{-3}$, assuming a Gaussian distribution. The
             integrated probability over the dark-shaded region amounts to
             5\%. In the absence of theoretical uncertainties, the light-shaded
             region is accessible in Model-II only for $M_{H^{\pm}} > 1276\,$GeV.
             Right: Confidence belts ($95\%\,$C.L.) in Model-II for the
             same experimental error, and including the theoretical
             uncertainties (see the text). The experimental central value from
             Eq.~(\ref{rexp}) is marked by the vertical dashed
             line. \label{fig:probdens}}
\end{center}
\end{figure}

Comparing the uncertainties in the four alternative averages for $R_\gamma$ at
$E_0=1.6\,$GeV in the last column of Tab.~\ref{tab:averages}, one can see that
the first two of them are less accurate. Thus, at the moment, the
balance of the background subtraction and extrapolation uncertainties points
towards using the results extrapolated from 1.9 or 2.0, at least when one
takes the errors from Ref.~\cite{Buchmuller:2005zv} for granted. Since there
is not much difference in the uncertainties of these two averages, we suggest
discarding the 2.0 one, as it requires a longer extrapolation. Thus, we
recommend adopting
\be \label{rexp}
R_{\gamma}^{\exp} = (3.22 \pm 0.15) \times 10^{-3}
\ee
as the current experimental average for $R_\gamma$ at $E_0=1.6\,$GeV.

\newsection{Bounds on $M_{H^\pm}$\label{sec:bounds}}

In this section, we shall use $R_\gamma$ to derive bounds on $M_{H^\pm}$ in
the 2HDM. We are going to treat all the uncertainties as stemming from
Gaussian probability distributions, which is obviously an ad-hoc assumption,
although consistent with combining various partial uncertainties in quadrature
on the theory side, and in the experimental averages. In any case, the quoted
confidence levels of our bounds should be taken with a grain of salt.

The left plot in Fig.~\ref{fig:probdens} shows a Gaussian probability
distribution for our average in Eq.~(\ref{rexp}). In Model-II, only
enhancements of $R_\gamma$ with respect to the SM prediction~(\ref{rsm}) are
possible. Thus, if there were no theoretical uncertainties, only $R_\gamma >
3.31 \times 10^{-3}$ would be accessible in Model-II. This is marked by the
shaded regions (both light and dark) in the considered plot. The
integrated probability over the dark-shaded region amounts to 5\%. The border
between the light- and dark-shaded regions corresponds to the central value for
$R_\gamma$ obtained for $M_{H^{\pm}} \simeq 1276\,$GeV in the limit $\cot\beta
\to 0$. Thus, one might expect that the $95\%\,$C.L. lower bound for
$M_{H^{\pm}}$ should amount to $1276\,$GeV in the absence of
theoretical uncertainties. We are not assuming here that Model-II is valid for
sure. Instead, we are allowing for a possibility that it gets excluded
(together with the SM) if $R_{\gamma}^{\exp}$ is sufficiently far below the SM
prediction.
\begin{table}[t]
\begin{center}
\begin{tabular}{|c|c|c|c|c|c|c|c|}\hline
Model & $R_\gamma^{\rm exp} \times 10^3$ 
& \multicolumn{3}{c|}{ $95\%\,$C.L. bounds } & \multicolumn{3}{c|}{ $99\%\,$C.L. bounds } \\
                &                 & 1-sided & 2-sided & FC & 1-sided & 2-sided & FC \\\hline
                & $3.05 \pm 0.28$ & 307 & 268 & 268 & 230 & 208 & 208 \\
      I         & $3.12 \pm 0.19$ & 401 & 356 & 356 & 313 & 288 & 288 \\
($\tan\beta=1$) & {\boldmath $3.22 \pm 0.15$} & {\bf 504} & {\bf 445} & {\bf 445} & {\bf 391} & {\bf 361} & {\bf 361} \\\hline
                & $3.05 \pm 0.28$ & 740 & 591 & 569 & 477 & 420 & 411 \\
      II        & $3.12 \pm 0.19$ & 795 & 645 & 628 & 528 & 468 & 461 \\
(absolute)      & {\boldmath $3.22 \pm 0.15$} & {\bf 692} & {\bf 583} & {\bf 580} & {\bf 490} & {\bf 440} & {\bf 439} \\\hline
\end{tabular}
\end{center}
\caption{\sf Bounds on $M_{H^{\pm}}$ obtained using different methods.\label{tab:bounds}}
\end{table}

To include the theory uncertainties, one follows the standard confidence belt
construction (see, e.g., Sec.~39.4.2.1 of Ref.~\cite{Olive:2016xmw}). For each
$M_{H^{\pm}}$, one considers a Gaussian probability distribution around the
{\em theoretical} central value, with its variance obtained by combining the
experimental and theoretical uncertainties in quadrature. Next, a confidence
interval corresponding to (say) $95\%$ integrated probability is
determined. It can be placed either centrally (for a derivation of 2-sided
bounds), or maximally shifted in either way (for 1-sided bounds), or in an
intermediate way, like in the Feldman-Cousins (FC)
approach~\cite{Feldman:1997qc}. This is illustrated in the right plot of
Fig.~\ref{fig:probdens}, for Model-II with $\cot\beta \to 0$. The
red, black, green, and blue intervals correspond to the 2-sided, upper
1-sided, lower 1-sided and FC cases, respectively.  The experimental central
value from Eq.~(\ref{rexp}) is marked by the vertical dashed line. On the
vertical axis, we use $1\,{\rm TeV}/M_{H^{\pm}}$ that is restricted to be
positive, which makes our case very similar to the one in Sec.~IV-B of
Ref.~\cite{Feldman:1997qc}.

It is the freedom of the confidence interval placement that makes the
resulting bounds on $M_{H^{\pm}}$ somewhat ambiguous.  If we choose the FC
(blue) intervals, low values of $R_{\gamma}^{\exp}$ can never lead to
exclusion of Model-II in its whole parameter space. If we choose the upper
1-sided (black) intervals, our method is actually equivalent to using the
experimental upper bound on $R_{\gamma}^{\exp}$ rather than the actual
measurement. In this case, the previously discussed example from the left plot
of Fig.~\ref{fig:probdens} is recovered in the limit of no theory
uncertainties. In the literature, bounds on $M_{H^{\pm}}$ have been derived
using either the 1-sided (e.g.,
Refs.~\cite{Misiak:2015xwa,Hermann:2012fc,Gambino:2001ew,Ciuchini:1997xe}) or
2-sided (e.g., Refs.~\cite{Belle:2016ufb,Flacher:2008zq}) approaches, and the
method choice was not always explicitly spelled out.

In Tab.~\ref{tab:bounds}, we present the bounds we obtain following three
different methods, and using three out of four\footnote{
We omit the one requiring the longest ($2.0 \to 1.6$) extrapolation in $E_0$.}
averages for $R_{\gamma}^{\exp}$ from Tab.~\ref{tab:averages}. The rows corresponding
to our preferred choice (Eq.~(\ref{rexp})) are displayed in bold. For Model-I we set
$\tan\beta = 1$, while the absolute bounds ($\cot\beta \to 0$) are shown for Model-II.
In the Model-I case, the lower rather than the upper 1-sided intervals are employed.

It is interesting to observe that stronger bounds on $M_{H^{\pm}}$ in Model-II
are found from the two less precise averages, just because their central
values turn out to be lower. These averages are less sensitive to the
$E_0$-extrapolation issues, which might be helpful in accepting the ones
derived from Eq.~(\ref{rexp}) as conservative. The situation in Model-I is
reverse -- the most precise average gives the strongest bounds, as naively
expected.

By coincidence, our 2-sided $95\%\,$C.L. bound of $583\,$GeV in Model-II
practically overlaps with the $580\,$GeV one that has been obtained in
Ref.~\cite{Belle:2016ufb} from their single measurement alone (giving
${\mathcal B}_{(s+d)\gamma}$ with a lower central value but larger uncertainty
than the one corresponding to our Eq.~(\ref{rexp})). Since this bound is also
the most conservative one, we suggest choosing it for updated combinations with
constraints from other observables.
\begin{figure}[t]
\begin{center}
\includegraphics[width=10cm,angle=0]{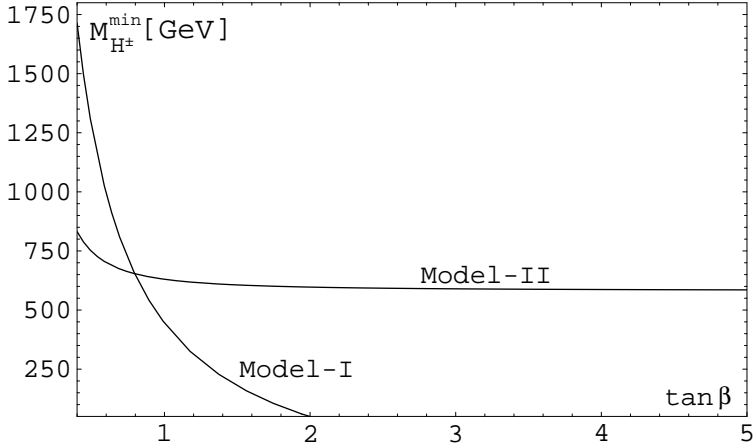}\hspace{5mm}
\caption{\sf $95\%\,$C.L. lower bounds on $M_{H^{\pm}}$ as functions of $\tan\beta$.\label{fig:tanb}}
\end{center}
\end{figure}

In Fig.~\ref{fig:tanb}, the $95\%\,$C.L. bounds on $M_{H^{\pm}}$ are shown
as functions of $\tan\beta$. Above $\tan\beta \simeq 2$, the Model-I bound 
becomes weaker than the LEP one ($\simeq 80\,$GeV~\cite{Olive:2016xmw}), while
the Model-II one gets saturated by its $\tan\beta \to \infty$ limit ($\simeq
580\,$GeV). Our plot terminates on the left side at $\tan\beta = 0.4$. For
lower values of $\tan\beta$, the bound from $R_b$ becomes more important in
Model-II (see Figs.~13 and 14 of Ref.~\cite{Flacher:2008zq}), while $R_\gamma$
alone in Model-I becomes insufficient due to possible changes of sign in the
coefficient $C_7$.  In the latter case, including the $b \to s \ell^+ \ell^-$
observables becomes necessary -- see Ref.~\cite{Blake:2016olu} and references
therein.

\newsection{Conclusions \label{sec:concl}}

We derived updated constraints on $M_{H^{\pm}}$ in the 2HDM that get imposed
by measurements of the inclusive weak radiative $B$-meson decay branching
ratio. Although in principle straightforward, such a derivation faces several
ambiguities stemming mainly from the photon energy cutoff choice. We presented
an extended discussion of this issue, and updated the experimental averages.
In Model-I, relevant constraints are obtained only for $\tan\beta \lsim
2$. In Model-II, the absolute ($\tan\beta$-independent)
$95\%\,$C.L. bounds are in the $570$--$800\,$GeV range. We recommend one of
the most conservative choices, namely $580\,$GeV, to be used for combinations
with constraints from other observables. This value overlaps with the bound
derived from the most recent single measurement alone~\cite{Belle:2016ufb}.

\section*{Acknowledgments}

We are grateful to Akimasa Ishikawa, Phillip Urquijo, S{\l}awomir~Tkaczyk
 and Aleksander~Filip \.Zarnecki for helpful discussions, as well as to Paolo
 Gambino for reading the manuscript and useful comments. The research of
 M.S. has been supported by the BMBF grant 05H15VKCCA. M.M. acknowledges
 partial support from the National Science Centre (Poland) research project,
 decision no.\ DEC-2014/13/B/ST2/03969, as well as by the Munich Institute for
 Astro- and Particle Physics (MIAPP) of the DFG cluster of excellence ``Origin
 and Structure of the Universe''.

\end{document}